# Optimal dispatch of low-carbon integrated energy system considering nuclear heating and carbon trading

Yang Li [a,*], Fanjin Bu[a], Jiankai Gao[a], Guoqing Li[a]

[a] School of Electrical Engineering, Northeast Electric Power University, Jilin 132012, China

* Corresponding author. E-mail address: liyang@neepu.edu.cn (Y. Li).

**Abstract:** The development of miniaturized nuclear power (NP) units and the improvement of the carbon trading market provide a new way to realize the low-carbon operation of integrated energy systems (IES). In this study, NP units and carbon trading mechanisms are introduced into the IES to build a new low-carbon scheduling model. In view of the decrease in system operation flexibility caused by the introduction of NP unit, on the one hand, the heating renovation of the NP unit is carried out to make it a cogeneration unit, to expand its operating range and improve its operating flexibility; on the other hand, auxiliary equipment such as electricity storage system, heat storage system and power to gas unit, which can carry out energy time translation or energy form conversion, are introduced into IES to jointly improve the flexibility of system operation. In the model-solving stage, the chance-constrained programming (CCP) model considering the uncertainty of the renewable energy (RE) output is converted into an equivalent mixed-integer linear programming (MILP) model using discretized step transformation. The test system built based on real data of an IES in North China shows that the proposed method has good economic and low-carbon environmental protection benefits.

**Key words:** Integrated energy system; Nuclear power unit; Carbon trading; Carbon emission; Nuclear heating; Low-carbon; Chance constraints; Renewable generation uncertainty.

## NOMENCLATURE

| | | | |
|---|---|---|---|
| Acronyms | | IES | Integrated energy system |
| NP | Nuclear power | ESS | Electricity storage system |
| HSS | Heat storage system | P2G | Power to gas |
| CCP | Chance-constrained programming | RE | Renewable energy |
| MILP | Mixed-integer linear programming | DST | Discretized step transformation |
| TP | Thermal power | GP | Gas power |
| PV | Photovoltaic | WT | Wind turbines |
| GC | Gas cogeneration | PDF | Probability density functions |
| HIA | Hybrid intelligent algorithm | IPM | Interior point method |
| Symbols | | $P_e^N$ | Electrical power of NP unit |
| $P_h^N$ | Heating power of NP unit | $P^N$ | Equivalent electrical power of NP |
| $P_{e,max}^N$ | Maximum electrical power of NP unit | $P_{e,min}^N$ | Minimum electrical power of NP unit |
| $V_{P2G}$ | Volume of natural gas from P2G | $P^{P2G}$ | Electrical power of P2G unit |
| $\eta^{P2G}$ | Efficiency factor of P2G unit | HHV | Calorific value of natural gas |
| $S_0$ | Initial energy of ESS | $S_{end}$ | Superfluous energy of ESS |
| $S_{min}$ | Minimum capacity of ESS | $S_{max}$ | Maximum capacity of ESS |
| $P_c^E$ | Charge power of HSS | $P_d^E$ | Discharge power of HSS |
| $C_0$ | Initial energy of ESS | $C_{end}$ | Superfluous energy of ESS |

| Symbol | Description | Symbol | Description |
|---|---|---|---|
| $P_e^{GC}$ | electrical power of GC unit | $P_h^{GC}$ | Heating power of GC unit |
| $P_{e,max}^{GC}$ | Maximum electrical power of GC unit | $P_{e,min}^{GC}$ | Maximum electrical power of GC unit |
| $P_{h,max}^{GC}$ | Maximum heating power of GC unit | $P_{h,min}^{GC}$ | Minimum heating power of GC unit |
| $V_{GC}$ | Natural gas consumed by GC unit | $Cg$ | Thermoelectric ratio of GC unit |
| $\eta_{loss}$ | Heat loss rate of GC unit | $E_f$ | Initial carbon emission quotas |
| $f$ | Coefficient of carbon quota | $Er$ | Actual total $CO_2$ emission |
| $k_c$ | Unit price of $CO_2$ emission right | $F$ | Operating cost of IES |
| $P^{th}$ | Electrical power of TP unit | $R^{th}$ | Spinning reserve of TP unit |
| $a$ | Cost coefficients of TP unit | $b$ | Cost coefficients of TP unit |
| $c$ | Cost coefficients of TP unit | $w$ | Spinning reserve cost coefficient |
| $\mu_{GC}$ | Price of natural gas | $R^{GC}$ | Spinning reserve of TP unit |
| $\delta$ | Spinning reserve cost coefficient | $g_1$ | Charge cost coefficients of ESS |
| $g_2$ | Discharge cost coefficients of ESS | $R^E$ | Spinning reserve of ESS |
| $\beta$ | Cost coefficients of NP unit | $P_r$ | Electrical power of RE |
| $P_l$ | Electrical load | $P_l^h$ | Heating load |
| $\alpha$ | Confidence level | $b(i_{b,t})$ | Probabilistic sequence of WT |
| $a(i_{a,t})$ | Probabilistic sequence of PV | $N_{a,t}$ | Sequence length of PV outputs |
| $l$ | Length of the discretization step | $c(i_{ct})$ | Probabilistic sequence of RE |
| $Zm_{ct}$ | 0-1 variable | $Q$ | A large positive number |
| $\lambda$ | A positive number close to 0 | | |

# 1 Introduction

Global warming caused by $CO_2$ emissions has become a major problem for human society. The electric power energy industry is an important source of $CO_2$ emissions and its low-carbon operation is of great significance for the sustainable development of society (Qu et al., 2018). An integrated energy system (IES) refers to a new type of energy system that uses advanced physical information technology and management mode in a certain area, comprehensively utilizes electric energy, heat energy, natural gas, and other energy sources in the area, and realizes coordinated planning and optimized operation among various energy subsystems (Li et al., 2021a). As an important carrier for utilizing wind power and photovoltaics, IESs have made outstanding contributions to the low-carbon operation of energy systems (Li et al., 2022a). However, for a large-scale IES, the thermal power (TP) and gas power (GP) units that undertake the basic load rely on the combustion of fossil fuels for energy supply, resulting in large amounts of $CO_2$ emissions. NP units rely on the nuclear reaction of nuclear fuel for energy supply and there is no $CO_2$ emission during operation, which makes NP units an important channel for promoting low-carbon societal development (Sui et al., 2018). In recent years, the miniaturization of NP plants has become increasingly mature, and several small NP plants have been constructed. The miniaturization of NP plants provides a good opportunity for integration with IES (HoM et al., 2019). However, owing to safety and economic considerations, reactor power cannot be changed frequently, which brings great challenges to the flexibility of the IES (Wang et al., 2018). Therefore, the introduction of NP plants into IES and the improvement of their operational flexibility are of great significance for energy transition.

Several studies have been conducted on the low-carbon operation of IES, and many constructive measures have been proposed based on existing research. The measures for low-carbon operation of IES are mainly of two types: policy measures, such as formulating corresponding reward and punishment policies for $CO_2$ emissions, and technical measures, such as $CO_2$ emission reduction or $CO_2$ capture through technological innovation.

The introduction of a carbon trading mechanism into an IES is a major policy measure for realizing low-carbon operations. A previous study (Xiao et al., 2018) introduced carbon trading into an electricity-gas IES, proving that carbon trading is conducive to the development of a low-carbon economy. The case of (Lin et al., 2019) indicates that the establishment of a carbon trading market is beneficial to the grid connection of clean units. Lu et al. (2021) put forward an operational model based on taking full advantage of the user demand response and carbon trading mechanism for low-carbon scheduling of IES. The results show that an appropriate carbon emissions trading mechanism can promote the low-carbon development of an IES. Although carbon trading among fossil fuel units can promote the priority operation of units with low carbon intensity, it inevitably generates significant carbon emissions. Therefore, its independent operation has a limited effect on the reduction in total carbon emissions. Technical measures include the introduction of carbon capture technology and oxygen-enriched combustion technology into the IES. Relevant studies have shown that the introduction of such technical measures significantly reduces IES carbon emissions (Yong et al., 2016). Although carbon capture technology and oxygen-enriched combustion technology can greatly reduce total carbon emissions, they require considerable energy in the working process, resulting in high energy consumption and low energy efficiency. In addition, such technical measures have some problems, such as high additional cost of operation and difficulty of transformation. Therefore, relevant research mainly focuses on the discussion of their operation mechanism and technical economy, and their application value is low.

The development of miniaturized NP units provides a new direction for the exploration of technical measures for low-carbon operation of IES. There is no $CO_2$ emission during the operation of the NP plant, which can reduce the total $CO_2$ emissions of the IES from the source. Kan et al. (2020) demonstrated the contribution of introducing NP units in the reduction in regional carbon emissions. Wang et al. (2017) constructed a model for NP units to participate in the peak adjustment of the power grid, considering the economic and low-carbon characteristics of NP units to a certain extent. However, this dispatch model has significant limitations concerning its application in actual operation. During the one-day dispatch period of an IES, frequent power changes of NP units have a negative effect on the stability of the nuclear reactor and the primary circuit (Li et al., 2014), and also bring about a decline in economics. Therefore, the NP unit should maintain a constant full-power operation mode within the one-day dispatch period, which poses a significant challenge to the flexibility of the IES operation. However, the uncertainty of the distributed RE output in the IES has high requirements for system flexibility.

It is proposed in literature (Garcia et al., 2013) that the excess heating energy of the NP unit reactor should be supplied to a nearby chemical plant to avoid frequent changes in the reactor power, which provides a new idea for improving the operational flexibility of the NP unit. Rinne et al. (2015) used the energy system in Finland as an example to discuss the improvement in the economy and operational flexibility of NP units by heating renovation. As stated in a previous study (Rämä et al., 2020), the heating renovation and thermoelectric characteristics of NP units are similar to those of conventional TP units. In addition, some pioneering studies have investigated the joint operation of NP units and auxiliary equipment. Power to gas (P2G) units can be started and stopped on demand without a significant impact on hydrogen quality and are therefore considered to be good containers for absorbing the remaining energy of the reactor (Choi et al., 2017). Research (Locatelli et al., 2018) has demonstrated the advantages of combining P2G and NP units. In the literature (Ruth et al., 2014), large-capacity batteries have been proposed as buffers for the energy output of reactors. Studies (Bose et al., 2017) show that a combined system of high-capacity electricity storage system (ESS) and NP units can better cope with external loads and renewable energy (RE) variations. Li et al. (2020) analyzed the IES with auxiliary equipment such as ESS, heat storage system (HSS), and P2G units and found that the energy storage system can realize the time transfer of energy, which greatly improves the flexibility of system operation.

Table 1 summarizes the main differences between the low-carbon IES scheduling method proposed in this study and the most relevant research studies in the field. In the Table 1, symbols "√" and "×" indicate if a particular feature is considered or not, respectively. The above studies have put forward many constructive measures for low-carbon operation of IES, but there are still some aspects worthy of further exploration. (1) Policy measures such as carbon trading mechanisms can promote the priority operation of clean units, but their effect on total $CO_2$ emissions is limited, so they need to be combined with other technical measures to better reduce emissions. (2) The technical measures for low-carbon transformation, such as carbon capture technology and eutrophic combustion technology, are mostly in the theoretical analysis stage owing to high investment and low overall thermal efficiency. (3) Although NP units have good performance in low-carbon operation, their unique operation mode brings great challenges to the operational flexibility of IES.

**Table 1. Comparison of the proposed low-carbon IES scheduling method with the most relevant studies**

| | Policy measure | Technical measure | Heating renovation | Auxiliary equipment | | | RE uncertainties | |
|---|---|---|---|---|---|---|---|---|
| | | | | P2G | ESS | HSS | PV | WT |
| Xiao et al., 2018 | √ | × | —— | × | × | × | √ | × |
| Lu et al., 2021 | √ | × | —— | × | √ | × | × | √ |
| Yong et al., 2016 | × | carbon capture | —— | × | √ | × | √ | √ |
| Wang et al., 2017 | × | NP unit | × | × | × | × | √ | √ |
| Garcia et al., 2013 | × | NP unit | √ | × | × | × | × | × |
| Rinne et al., 2015 | × | NP unit | √ | × | × | × | × | √ |
| Choi et al., 2017 | × | NP unit | × | √ | × | × | √ | √ |
| Ruth et al., 2014 | × | NP unit | × | × | √ | × | √ | √ |
| This paper | √ | NP unit | √ | √ | √ | √ | √ | √ |

In response to the above problems, this study actively explores the low-carbon operation of an IES. The main contributions of this study are summarized as follows:

(1) To reduce the overall carbon emissions, introducing small nuclear power unit to the IES can provide a stable energy supply without producing CO2 during operation. We then design a carbon trading mechanism based on a stepped carbon price model, which translates the low-carbon characteristics of NP units into economic advantages and promotes the low-carbon development of the IES.

(2) To deal with the lack of system operational flexibility caused by the integration of NP units, the NP units are transformed into cogeneration units by heating renovation to expand their operation range. Additionally, auxiliary equipment such as ESS, HSS, and P2G unit further improve the flexibility of system operation by energy time shift or form conversion.

(3) To handle the uncertainty of the distributed RE output, a low-carbon IES scheduling model

with chance constraints is built, and a new discretized step transformation (DST) method is employed to handle chance constraints. Using the probabilistic sequences of the RE output distribution, the chance constraint can be directly transformed into its deterministic equivalent constraint without the need for the inverse function of the distribution, which is a novel method for solving CCP issues.

(4) Extensive tests were performed to examine the effectiveness and superiority of the proposed method using a real-world IES located in northern China. In particular, an analysis of carbon emissions and operating costs of the IES was conducted.

## 2 Low-carbon IES model
### 2.1 System structure

A low-carbon IES that realizes the comprehensive utilization of nuclear energy is shown in Fig. 1. Low-carbon IES mainly include photovoltaic (PV) and wind turbines (WT), TP units, small NP units, gas cogeneration (GC) units, P2G units, energy storage systems, and other auxiliary equipment.

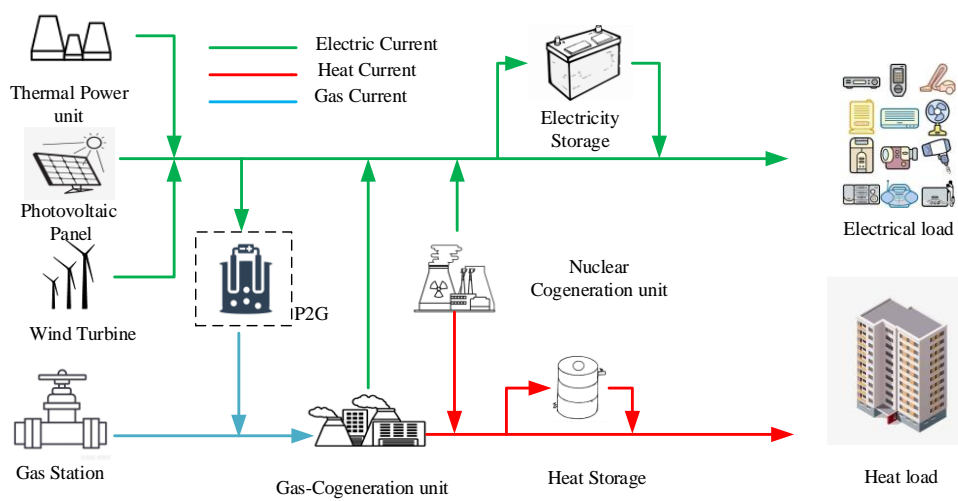

Fig. 1 System structure diagram

### 2.2 Nuclear cogeneration unit model

The miniaturization of NP units has attracted the attention of all countries worldwide to adapt to the many new characteristics of power systems. Since the beginning of the new century, the process of miniaturizing NP units has achieved breakthrough development, and a variety of small nuclear reactors have been developed, such as Russia's KLT. -40 as well as the Chinese ACP-100. In this study, an ACP100 unit was used as a prototype for the model. Although NP units have a certain peak-shaving capability in design, based on safety and economic considerations, in practice, NP units do not participate in daily peak shaving and operate at a constant full power state within the one-day dispatch period, which brings enormous challenges to the flexibility of the operation of IES.

1) Safety

Safety is the most basic and important requirement of NP units. Operational data show that frequent power changes in NP units reduce the service life of control rods and the reliability of primary circuit equipment, which is not conducive to the stable control of nuclear reactors and increases the risk of human error (Preischl et al., 2013).

2) Economical

The NP units adopt the operational mode of replacing nuclear fuel in a fixed cycle; therefore, reducing the power during the operation cycle causes the waste of nuclear fuel and increases the cost of spent fuel processing (Zhao et al., 2020). In addition, the power regulation of nuclear reactors can lead to a substantial increase in electricity consumption within a power plant.

As shown in Fig. 2, the electrical power of the NP unit is constant at point A during operation, and its flexibility is poor.

$$P_{e,t}^N = P_{e,\max}^N \quad \forall t \tag{1}$$

where $P_{e,\max}^N$ denotes the maximum electrical power of the NP unit. This is an effective measure to improve the system flexibility to transform it into a cogeneration unit.

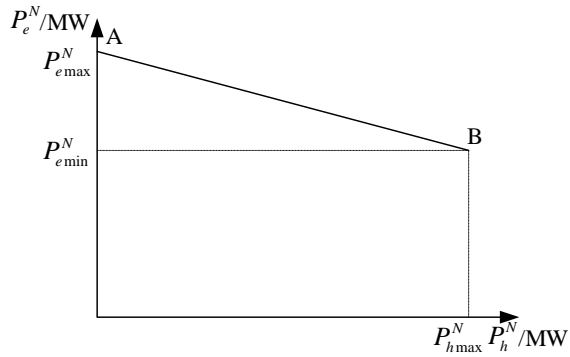

Fig. 2 Thermoelectric characteristics of nuclear cogeneration unit

There are two main types of heating renovations for NP units. One is to use a recovery device to recover the waste heat of steam after power generation for heating. This method does not cause the loss of power generation and is only suitable for supplying heat to a small area that is close to the power plant. The other is to pump steam from the steam turbine for heating. The principle is the same as that of conventional cogeneration units (Rämä et al., 2020), as shown in Fig. 3.

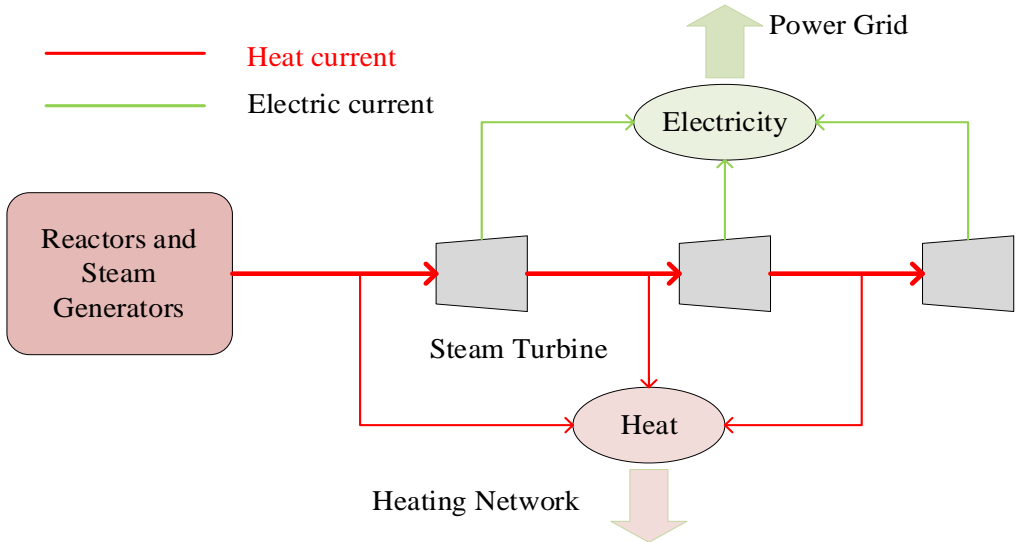

Fig. 3 Schematic diagram of the nuclear cogeneration unit

In this study, the second method is adopted. Pumping steam from the steam turbine for heating supply inevitably leads to a loss of power generation. Thus, the heating power $P_h^N$, electric power $P_e^N$, and equivalent electric power $P^N$ under pure condensation conditions satisfy the following relationship:

$$P_{it}^N = P_{e,it}^N + C_{V,i} P_{h,it}^N \tag{2}$$

$$P_{it}^N = P_{e,\max}^N \qquad \forall t \tag{3}$$

where $C_{V,i}$ denotes the thermoelectric ratio of the cogeneration unit. As shown in Fig. 2, after heating renovation, the operating range of the NP unit was expanded from point A to line segment A-B, which significantly enhanced the flexibility of its operation.

In addition, because the equivalent electric power $P^N$ is guaranteed to remain unchanged, the reactor power is in a constant full-power operating state, which ensures the stability of the primary circuit condition. It should be noted that the operating range of the NP cogeneration unit is line segment A-B rather than the irregular operating area of the ordinary cogeneration unit.

### 2.3 P2G unit model

P2G converts electrical energy into gaseous fuel. Specifically, it uses electrical energy to electrolyze water into oxygen and hydrogen and then further converts hydrogen into methane. The P2G equipment opens the channel for converting electrical energy into chemical energy of natural gas, realizes bidirectional conversion between electric and natural gas systems, expands the energy circulation path, and improves the flexibility of the IES. The P2G unit has a wide range of application scenarios to alleviate the time mismatch between source and load demands. When the electrical load is low, the P2G unit can convert excess electrical energy into chemical energy and store it in a natural-gas system. At the same time, the capture and utilization of $CO_2$ is realized in the P2G process, which has good low-carbon environmental benefits. The P2G process involves the following two chemical reactions (Li et al., 2022b):

$$\begin{cases} 2H_2O \rightarrow 2H_2 + O_2 \\ 4H_2 + CO_2 \rightarrow CH_4 + 2H_2O \end{cases} \tag{4}$$

The energy conversion of P2G unit can be described by the following formula:

$$V_{P2G} = \frac{\eta^{P2G} \cdot P_t^{P2G} \cdot \Delta t \cdot \varepsilon}{HHV} \tag{5}$$

where $V_{P2G}$ is the volume of natural gas, $P_t^{P2G}$ is the electrical power, $\eta^{P2G}$ is the efficiency factor of the P2G unit, $\varepsilon$ is the energy conversion coefficient, and $HHV$ is the calorific value of the natural gas per unit volume.

### 2.4 Energy storage system model

With a reasonable energy storage and release strategy, the energy storage device can suppress the fluctuation caused by variations in RE output and load. This study considered two types of energy storage devices: ESS and HSS.

#### 2.4.1 Electric energy storage system model

The ESS can reasonably realize the time decoupling of electric energy production and load demand. When the load level is relatively low or the RE output is high, the ESS stores electrical energy. When the electric load is high, the ESS releases electrical energy, which greatly alleviates the time mismatch between electrical energy production and consumption and increases the flexibility of the system. The ESS energy storage in each period can be expressed by the following formula (Li et al., 2019):

$$\begin{cases} S_{t+1} = S_t + \eta^E \left( P_{c,t}^E - P_{d,t}^E \right) \cdot \Delta t \\ S_0 = S_{end} \end{cases} \quad (6)$$

where $\eta^E$ is the charge and discharge efficiency; $P_c^E$ and $P_d^E$ are the charge and discharge power, respectively; $S_0$ is the battery capacity at the beginning of the dispatching time; and $S_{end}$ is the battery capacity at the end of the dispatching time.

**2.4.2 Heat storage system model**

The combined operation of the HSS and cogeneration units can decouple the thermoelectric connection of the cogeneration unit and improve the flexibility of the IES. When the electricity demand is low, the cogeneration unit reduces the electrical power to improve the heating power and store the heat energy in the HSS. When the electricity demand is high, part of the heating demand is undertaken by the HSS, so the cogeneration unit can further reduce the heating power and increase the electrical power to meet the electrical demand. The mathematical model of the heat storage capacity is (Pan et al., 2010):

$$\begin{cases} C_t = C_{t-1} - P_{h,t}^c \cdot \Delta t \\ C_0 = C_{end} \end{cases} \quad (7)$$

where $C_t$ is the heat storage capacity of HSS in the period $t$; $P_{h,t}^c$ is the storage and release power of HSS, which is positive at the time of heat storage and negative at the time of heat release. $C_0$ is the heat stored by the HSS at the beginning of the scheduling time and $C_{end}$ is the heat stored by the HSS at the end of the scheduling time.

**2.5 Gas cogeneration unit model**

Compared with coal-fired combined heat and power (CHP) unit, GC units have the advantages of high thermal efficiency and a small footprint. Under the same power generation conditions, the GP unit releases less $CO_2$, which has attracted increasing attention and development. The thermoelectric characteristics of the GC units are the same as those of ordinary CHP units (Roy P K et al., 2014), as shown in Fig. 4.

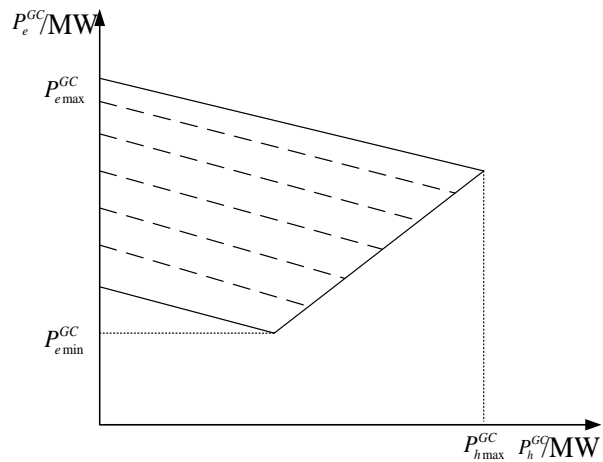

**Fig. 4 Thermoelectric characteristics of GC unit**

The shaded area represents the operating domain of the GC unit, where $P_{h\min}^{GC}$ and $P_h^{GC}$ represent the electrical power and heating power, respectively. $P_{e\min}^{GC}$, $P_{e\max}^{GC}$, $P_{h\max}^{GC}$, and $P_{h\min}^{GC}$ are the limits of electrical power and heating power, respectively. The volume of natural gas $V_{GC}$ consumed by the GC unit can be calculated using the following formula:

$$V_{GC} = \frac{\left[\left(P_e^{GC}/C_g\right) + P_h^{GC}\right]\Delta t \cdot \varepsilon}{(1-\eta_{loss})HHV} \tag{8}$$

where $C_g$ is the thermoelectric ratio of the GC unit and $\eta_{loss}$ is the heat loss rate.

### 2.6 Probabilistic model of photovoltaic and wind turbine output

Wind and solar power generation is the main form of RE utilization at present, and it has received extensive attention because of its many advantages, such as no pollution, no fossil energy consumption, and no CO₂ emissions. Photovoltaic and wind turbine power can be described by the probability density functions (PDF) of solar irradiance and wind speed, respectively. Some studies have indicated that wind speed and solar irradiance obey the Weibull and beta distributions, respectively. On this basis, our previous work carried out detailed research and derivation on the probabilistic models of photovoltaic (Li et al., 2019) and wind turbine output (Li et al., 2021b), which will not be repeated here.

### 2.7 Carbon trading model

The construction of the carbon trading market and carbon trading mechanism is an important policy measure. The carbon emission trading mechanism promotes the low-carbon operation of IES from a market perspective, and rewards and punishes carbon emission entities in the form of carbon emission trading rights. The initial carbon emission quotas are allocated through the baseline method, and the initial carbon emission quota of an IES is (Zhu et al., 2020):

$$E_f = f \cdot \sum_N \sum_T P_{n,t} \tag{9}$$

where $P_{n,t}$ is the equivalent power load of all energy demands at moment $t$ in community $n$, and $f$ is the coefficient of the carbon quota for free, which can be adjusted according to the emission reduction target. If the system's carbon emissions exceed the free limit, additional carbon emission rights need to be purchased; if the carbon emission is lower than the free limit, it can be sold in the carbon trading market for profit, so the carbon emission cost is (Wang et al., 2022):

$$T_c = k_c \cdot E \tag{10}$$

$$E = E_r - E_f \tag{11}$$

where $k_c$ is the unit price of CO₂ emission rights and $E_r$ is the actual total CO₂ emission. Unlike the traditional carbon trading mechanism with fixed carbon prices, this paper constructs a stepped carbon price model. The key idea behind this model is that with the increase of carbon emissions, the carbon trading price rises in steps, which will lead to higher operating costs and promote the low-carbon development of the IES. The specific stepped carbon price model is as follows:

$$k_c = \begin{cases} k_1 & E < E_1 \\ k_2 & E_1 \leq E \leq E_2 \\ k_3 & E > E_2 \end{cases} \quad (12)$$

Among them, $k_1$, $k_2$, and $k_3$ are the different carbon trading prices; $E_1$, $E_2$ are the carbon emission thresholds, which are determined by the overall load of the IES. The change of carbon trading price $k_c$ with carbon emissions $E$ is shown in Fig. 5.

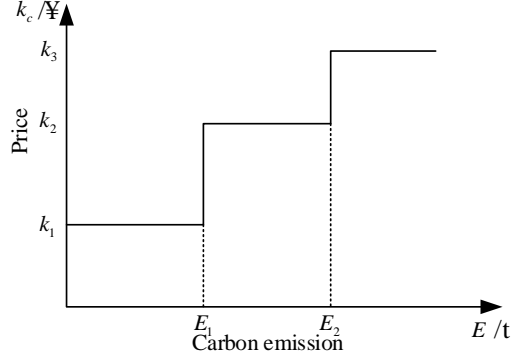

Fig. 5 Changes in carbon trading prices with carbon emissions

In this study, $E_r$ mainly considers carbon emissions from the operation of TP units $E^{th}$, GC units $E^{NG}$, and the absorption of $CO_2$ by the P2G unit $E_{P2G}$, their specific relationships are as follows:

$$E_r = E^{th} + E^{NG} - E_{P2G} \quad (13)$$

The carbon emissions of TP units are described by the following formula

$$E^{th} = \sum_{i=1}^{N}\sum_{t=1}^{T} B_i^{th} \cdot P_{it}^{th} \quad (14)$$

where $P_{it}^{th}$ is the electric power of the TP unit $i$ in time period $t$ and $B^{th}$ is the $CO_2$ emission intensity of the TP unit. Carbon emissions of the GC unit were calculated using the following formula:

$$E^{NG} = V_{GC} \cdot B^{NG} \quad (15)$$

where $B^{NG}$ is the amount of $CO_2$ produced by the combustion of natural gas per unit volume. The amount of $CO_2$ absorbed by the P2G unit is:

$$E_{P2G} = V_{P2G} \cdot B^{NG} \quad (16)$$

The basic principles of the carbon trading model constructed in this study are illustrated in Fig. 6.

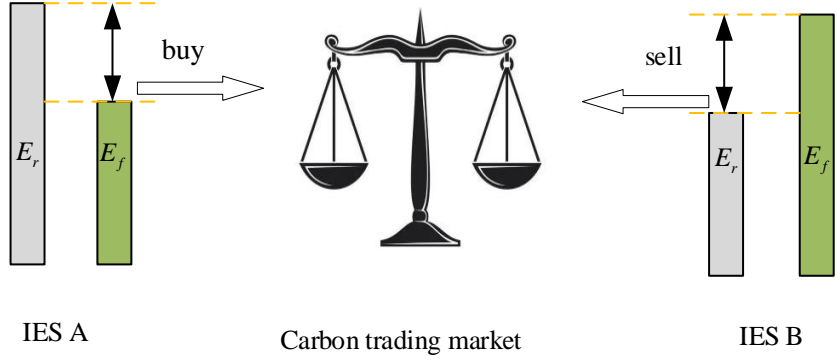

IES A      Carbon trading market      IES B

Fig. 6 Principle of carbon trading

Fig. 6 shows that the environmental benefits of carbon emission reduction can be transformed into economic benefits through the carbon trading mechanism, which is conducive to the priority operation of low-carbon emission units.

## 3 Scheduling Model

In this section, a scheduling model is constructed to determine the lowest operating cost and $CO_2$ emissions of an IES. The specific model is described in detail below:

### 3.1 Objective function

The operating cost of the IES includes the carbon trading and equipment operating costs, as shown below:

$$\min F = C_1\left(P_{it}^{th}\right) + C_2\left(P_{e,it}^{gc}, P_{h,it}^{gc}\right) + C_3\left(P_{d,t}^{E}, P_{c,t}^{E}\right) + C_4(P_{it}^{N}) + C_5(E) \quad (17)$$

where $F$ is the total cost of the IES, $C_1$ is the operating cost of the TP units, $C_2$ is the operating cost of the GC units, $C_3$ is the charging and discharging cost of the ESS, $C_4$ is the operating cost of the NP units, and $C_5$ is the carbon trading cost.

$$C_1 = \sum_{t=1}^{T}\sum_{i=1}^{n}\left[a_i\left(P_{it}^{th}\right)^2 + b_i P_{it}^{th} + c_i + w_i R_{it}^{th}\right] \quad (18)$$

where $P_{it}^{th}$ is the electric power of the TP unit $i$ in time period $t$, $R_{it}^{th}$ is the spinning reserve power of the TP unit $i$ in time period $t$, $a_i$, $b_i$ and $c_i$ are the fuel cost coefficients of the TP unit $i$, and $w_i$ is the spinning reserve cost coefficient.

$$C_2 = \mu_{GC}\cdot(V_{GC} - V_{P2G}) + \sum_{i=1}^{N}\sum_{t=1}^{t}\delta_i R_{it}^{GC} \quad (19)$$

where $\mu_{GC}$ is the price of natural gas per unit volume; $R_{it}^{GC}$ is the spinning reserve power provided by the GC unit $i$ at time $t$, and $\delta_i$ is the cost coefficient of the spinning reserve provided by the GC unit $i$.

$$C_3 = \sum_{t=1}^{T}\left[ g_1 P_{d,t}^{E} + g_2 P_{d,t}^{E} + \lambda R_t^{E} \right] \tag{20}$$

where $g_1$ and $g_2$ are the cost coefficients of the ESS charge and discharge, respectively; $\lambda$ is the cost coefficient of the ESS providing the spinning reserve, and $R_t^{E}$ is the spinning reserve provided by the ESS at time $t$.

$$C_4 = \sum_{i=1}^{m}\sum_{t=1}^{T} \beta \cdot P_{it}^{N} \tag{21}$$

where $\beta$ is the fuel cost coefficient of the NP unit and $P_{it}^{N}$ is the power of the NP unit. Because the NP unit maintains full power after it is put into operation and nuclear fuel is replaced in a fixed period, $\beta$ and $P_{it}^{N}$ are both fixed constants; therefore, the operating cost of the NP unit $C_4$ is also a fixed constant.

$$\begin{cases} C_5 = k_c \cdot E \\ E = E_r - E_f \end{cases} \tag{22}$$

where $E$ is the equivalent $CO_2$ emission.

### 3.2 Constraints
#### 3.2.1 Energy balance constraints

The energy balance constraints of the IES include electric power balance, heating power balance, and natural gas balance constraints:

$$\sum_{i=1}^{n} P_{it}^{th} + \sum_{i=1}^{N} P_{e,it}^{GC} + \eta^{E}\left(P_{d,t}^{E} - P_{c,t}^{E}\right) + \sum_{i=1}^{m} P_{e,it}^{N} + P_{r,t} = P_{l,t} + P_t^{P2G} \tag{23}$$

$$\sum_{i=1}^{n} P_{h,i}^{GC} + \sum_{i=1}^{m} P_{h,i}^{N} + P_h^c = p_{lt}^h \tag{24}$$

$$V_{GC} - V_{P2G} = V_p \tag{25}$$

where $P_r$ is the absorbed RE power, which is the sum of the absorbed photovoltaic and wind turbine outputs; $P_l$ and $P_l^h$ are the electrical and heating loads of the IES, respectively; and $V_p$ is the volume of natural gas purchased from the gas grid.

#### 3.2.2 Equipment operation constraints

(1) Thermal power unit constraints

The power of a TP unit must operate within allowable upper and lower limits.

$$P_{i\min} \leq P_{it}^{th} \leq P_{i\max} \tag{26}$$

where $P_{i\min}$ and $P_{i\max}$ are the upper and lower limits of the TP unit electric power, respectively. In addition, the power change rate of the TP unit should satisfy the following requirements:

$$-r_{di} \leq P_{it}^{th} - P_{i(t-1)}^{th} \leq r_{ui} \tag{27}$$

where $r_{di}$ and $r_{ui}$ are the maximum rates of increase and decrease, respectively, in the power of the TP unit.

(2) Gas cogeneration unit constraints:

The electric power and heating power of the GC unit should be maintained within a certain range.

$$\begin{cases} P_{e,i\min}^{GC} \leq P_{e,it}^{GC} \leq P_{e,i\max}^{GC} \\ 0 \leq P_{h,it}^{GC} \leq P_{h,i\max}^{GC} \end{cases} \quad (28)$$

where $P_{e,i\max}^{GC}$ and $P_{e,i\min}^{GC}$ are the upper and lower limits of the electric power of the GC unit, respectively, and $P_{h,i\max}^{GC}$ is the maximum heating power of the GC unit. In addition, the rates of power rise and fall of the GC unit should satisfy the following constraints:

$$-r_{di}^{GC} \leq P_{e,it}^{CG} - P_{e,i(t-1)}^{CG} \leq -r_{ui}^{GC} \quad (29)$$

where $r_{ui}^{GC}$ and $r_{di}^{GC}$ are the upper and lower limits, respectively, of the power rise and fall rates of the GC unit.

(3) Nuclear cogeneration unit constraints:

After heating the renovation of the NP unit, its operating range is greatly increased, but certain limiting conditions should still be met, specifically:

$$\begin{cases} P_{e,i\min}^{N} \leq P_{e,it}^{N} \leq P_{e,i\max}^{N} \\ 0 \leq P_{h,it}^{N} \leq P_{h,i\max}^{N} \end{cases} \quad (30)$$

where $P_{e,i\max}^{N}$ and $P_{e,i\min}^{N}$ are the upper and lower limits of the electric power, respectively, and $P_{h,i\max}^{N}$ is the upper limit of the heating power of the NC units.

(4) P2G unit constraints

The power of the P2G unit should be within its allowable range, and the rate of power rise and fall should also be within a certain range, with specific constraints as follows:

$$\begin{cases} 0 \leq P_t^{P2G} \leq P_{\max}^{P2G} \\ -r_{di}^{P2G} \leq P_t^{P2G} - P_{(t-1)}^{P2G} \leq r_{ui}^{P2G} \end{cases} \quad (31)$$

where $P_{\max}^{P2G}$ is the upper limit of P2G the operating power, $r_u^{P2G}$ and $r_d^{P2G}$ are the maximum rates of the P2G unit power rise and fall, respectively.

(5) Electricity storage system constraints:

The charging and discharging power of ESS must meet the following requirements:

$$\begin{cases} 0 \leq P_D^E \leq P_{D,\max}^E \\ 0 \leq P_C^E \leq P_{C,\max}^E \end{cases} \quad (32)$$

where $P_{D,\max}^E$ and $P_{C,\max}^E$ are the upper limits of the ESS charge and discharge powers, respectively. To protect the ESS and prolong its service life, its capacity should be maintained at a minimum value. Therefore, the following restriction is imposed.

$$S_{\min} \leq S_t \tag{33}$$

where $S_{\min}$ is the minimum capacity that the ESS should hold.

(6) Heat storage system constraints

Heat storage and release rate of HSS meet the following constraints (Bose et al., 2017):

$$-P_{h,\max}^c \leq P_h^c \leq P_{h,\max}^c \tag{34}$$

where $P_{h,\max}^c$ is maximum rate of heat storage and release of HSS.

(7) Spinning reserve constraint

TP unit spinning reserve constraints:

$$P_i^{th} + R_i^{th} \leq P_{i\max} \tag{35}$$

GC unit spinning constraints:

$$P_{e,i}^{GC} + R_i^{GC} \leq P_{e,i\max}^{GC} \tag{36}$$

ESS spinning reserve constraints:

$$R^e \leq \min\left[(S_t - S_{\min})/\Delta t, P_{D,\max}^E - P_{D,t}^E\right] \tag{37}$$

We obtained the expected output $E_t$ of RE from the probability model above. However, the actual output value of RE units $P_R$ is a random variable, so $P_R$ and $E_t$ may not be equal in actual operation. The spinning reserve provided by the system should be able to fill this possible gap. In extreme cases, the actual output of RE could be 0, but the probability of such a situation is extremely small, and the required spinning reserve capacity and the cost are extremely high. If the spinning reserve capacity is set according to such a situation, it will cause a huge resource cost (Li et al., 2018a). Therefore, we describe the constraints of the spinning reserve in the form of chance constraints, that is, the probability that the spinning reserve capacity is greater than the possible gap is greater than a certain confidence level, so that the balance of reliability and economy of the system can be achieved. The specific spinning reserve constraints are as follows:

$$p_r\left\{\sum_{i=1}^n R_{it}^{th} + \sum_{i=1}^N R_{e,it}^{GC} + R_t^e \geq E_t - P_{Rt}\right\} \geq \alpha \tag{38}$$

where $\alpha$ is the confidence level.

## 4 Model conversion and solution

The scheduling model constructed above contains chance constraints that are difficult to solve. We use the DST to transform these chance constraints into equivalent deterministic constraints and then linearize them to transform the CCP model into an MILP model that is easy to solve.

### 4.1 Model conversion
#### 4.1.1 Probabilistic sequence of RE output

Photovoltaic power generation and wind power generation in period $t$ are described by a probability density function, so probabilistic sequences $a(i_{at})$ and $b(i_{bt})$ can be obtained by discretization through sequence operation theory, and the length $N_{at}$ of the probability sequence of wind power generation is (Li et al., 2020):

$$N_{at} = [P_{WT\max t} / l] \qquad (39)$$

where [] is the ceiling function, $P_{WT\max t}$ is the maximum possible output of the WT in time $t$, and $l$ is the length of the discretization step. The WT output in state $m_a$ is $m_a l (0 \leq m_a \leq N_{at})$, so WT output and its probabilistic sequence $a(i_{at})$ can be seen in Table 2 (Li et al., 2020):

**Table 2. Probabilistic sequence of fan output**

| Output/MW | O | $l$ | ⋯ | $m_a l$ | ⋯ | $N_{at} l$ |
|---|---|---|---|---|---|---|
| Probabilistic | $a(0)$ | $a(1)$ | ⋯ | $a(m_a)$ | ⋯ | $a(N_{at})$ |

The probability sequences corresponding to different wind turbine output states can be calculated using the probability density function of the wind turbine output (Li et al., 2022b):

$$a(i_{at}) = \begin{cases} \int_0^{l/2} f_p(P_{WT}) dP_{WT}, & i_{at} = 0 \\ \int_{i_{at}l-l/2}^{i_{at}l+l/2} f_p(P_{WT}) dP_{WT}, & i_{at} > 0, i_{at} \neq N_{at} \\ \int_{i_{at}l-l/2}^{i_{at}l} f_p(P_{WT}) dP_{WT}, & i_{at} = N_{at} \end{cases} \qquad (40)$$

where $f_p(P_{WT})$ is the PDF of the wind-turbine output. Using the same method, we can obtain the photovoltaic output and its corresponding probabilistic sequence $b(i_{bt})$. The probabilistic sequence $c(i_{ct})$ corresponding to the common force of renewable output in time period $t$ can be obtained using the rolling sum of the probabilistic sequences $a(i_{at})$ and $b(i_{bt})$:

$$c(i_{ct}) = \sum_{i_{at}+i_{bt}=i_{ct}} a(i_{at}) b(i_{bt}), \quad i_{ct} = 0, 1, ..., N_{at} + N_{bt} \qquad (41)$$

As shown in Table 3, the RE output is a sequence with $l$ as the discretization step and $N_{ct}(N_{ct} = N_{at} + N_{bt})$ as the length. Each output value corresponds to a probability, and these probability values constitute the probabilistic sequence of RE output $c(i_{ct})$:

**Table 3. Probabilistic sequence of RE output**

| Output/MW | O | $l$ | $2l$ | ⋯ | $(N_{ct}-1)l$ | $N_{ct} l$ |
|---|---|---|---|---|---|---|
| Probabilistic | $c(0)$ | $c(1)$ | $c(2)$ | ⋯ | $c(N_{ct}-1)$ | $c(N_{ct})$ |

### 4.1.2 Transformation of chance constraints into deterministic constraints

To perform deterministic transformation, a new 0-1 variable $z_{m_{ct}}$ that satisfies the following relationship is defined:

$$\begin{cases} z_{m_{ct}} = \begin{cases} 1, & \sum_{i=1}^{n} R_{it} + \sum_{i=1}^{N} R_{e,it} + P_{\text{Re}sst} \geq E_t - m_{ct} l \\ 0, & otherwise \end{cases} \\ m_{ct} = 0, 1, ..., N_{at} + N_{bt} \end{cases} \qquad (42)$$

The above formula indicates that in period $t$, when the spinning reserve capacity of the system is greater than the difference between the expected output $E_t$ of RE and the output $m_{ct} l$ of RE of type $m_{ct}$ in Table 3, the value of $z_{m_{ct}}$ is 1; otherwise, the value is 0. As shown in Table 3, the spinning reserve constraint in this scheduling model can be expressed as:

$$\sum_{m_{ct}=0}^{N_{ct}} z_{m_{ct}} c(m_{ct}) \geq \alpha \tag{43}$$

**4.1.3 Linearization**

The expression $z_{m_{ct}}$ above cannot be solved by the mixed-integer linear programming method; therefore, the following linearization must be performed (Li et al., 2021a):

$$\left(\sum_{i=1}^{n} R_{it}^{th} + \sum_{i=1}^{N} R_{e,it}^{GC} + R_{t}^{e} - E_{t} + P_{Rt}\right)/Q \leq Z_{m_{ct}} \leq 1 + \left(\sum_{i=1}^{n} R_{it}^{th} + \sum_{i=1}^{N} R_{e,it}^{GC} + R_{t}^{e} - E_{t} + P_{Rt}\right)/Q \tag{44}$$

where $Q$ is a large positive number, and when $\sum_{i=1}^{n} R_{it}^{th} + \sum_{i=1}^{N} R_{e,it}^{GC} + R_{t}^{e} \geq E_{t} - P_{Rt}$, the above equation is equivalent to $\lambda \leq z_{m_{ct}} \leq 1 + \lambda$ ($\lambda$ is a positive number close to 0). Since $z_{m_{ct}}$ is a 0-1 variable, it can only be equal to 1; it can only be equal to 0 in the other case. Therefore, this expression has the same meaning as Equation (38). At this point, the CCP model is transformed into an MILP model using DST.

**4.2 Solution process**

The process of the proposed method is illustrated in Fig. 7. It mainly comprises the following steps.

Step1: Construct the thermoelectric characteristic model of NC unit after heating renovation;

Step2: Establish a reasonable carbon trading mechanism;

Step3: Construct the structure model of low-carbon IES;

Step4: Build an optimal scheduling model for a low-carbon IES with opportunity constraints.

Step5: Discretize the PDF of the RE output to generate the corresponding probability sequence.

Step6: Obtain the expected value of the scenery output in each period based on the DST

Step7: Transform the opportunity constraint of spinning reserve into the equivalent deterministic constraint;

Step8: Sort out the MILP scheduling model;

Step9: Input the original system parameters;

Step10: Use CPLEX solver to find the optimization scheduling scheme;

Step11: Check whether the scheme has been found. If yes, the program is terminated; otherwise, update the value of $\alpha$ and proceed to Step 9;

Step12: Apply the optimal scheduling scheme obtained to the low-carbon operation of the IES.

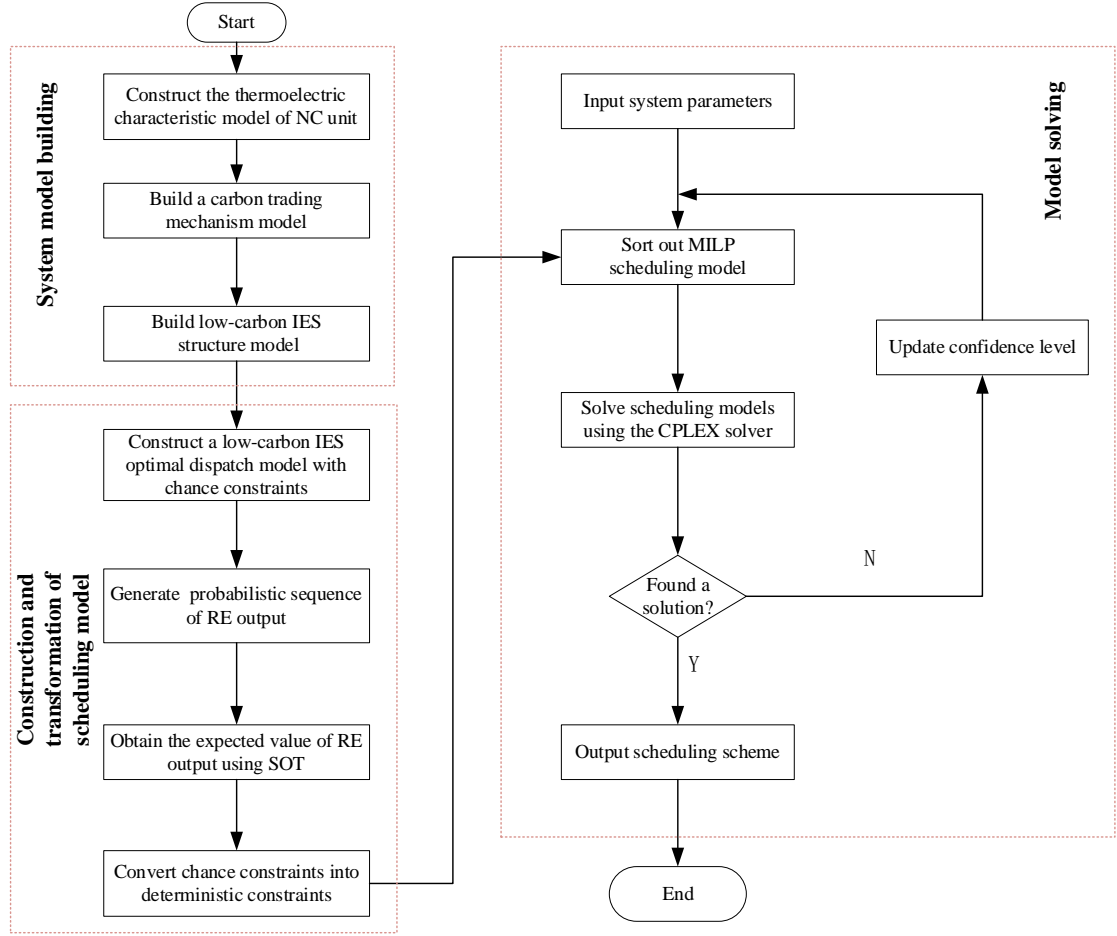

Fig. 7 Flow chart of the proposed method

## 5 Case study

To verify the advantages of the proposed method in the low-carbon economic operation of an IES, a test system was established using real data from a community in North China. All simulations and tests were conducted on a computer with a dual-core Intel CPU and 16 GB of RAM.

### 5.1 Test system overview

The test system of the electric-thermal-gas IES constructed in this study is shown in Fig. 1, in which the main equipment includes four TP units, two GC and NP units, one ESS, one HSS, one P2G unit, a distributed wind turbine, and photovoltaic units. The parameters of the TP and GC units are listed in Tables 4 and 5 (Li et al., 2020).

Table 4. Parameters of thermal power units

| $P_{i\max}$ | $P_{i\min}$ | $r_{di}$ | $r_{ui}$ | $a_i$ | $b_i$ | $c_i$ | $w_i$ | $B^{th}$ |
|---|---|---|---|---|---|---|---|---|
| MW | MW | MW/h | MW/h | ¥/MW$^2$ | ¥/MW | ¥ | ¥/MW | t/MWh |
| 40 | 12 | 20 | 20 | 0.18 | 237.25 | 113.02 | 40 | 0.97 |
| 40 | 12 | 20 | 20 | 0.18 | 237.25 | 113.02 | 40 | 0.97 |
| 30 | 10 | 15 | 15 | 0.13 | 245.31 | 99.52 | 38 | 1.06 |
| 30 | 10 | 15 | 15 | 0.13 | 245.31 | 99.52 | 38 | 1.06 |

Table 5. Parameters of gas cogeneration units

| $P_{e,i\max}^{GC}$ | $P_{e,i\min}^{GC}$ | $P_{h,i\max}^{GC}$ | $r_{di}^{GC}$ | $r_{ui}^{GC}$ | $\delta_i$ | $\eta_{loss}$ | $C_g$ | $\mu_{GC}$ | $B^{NG}$ |
|---|---|---|---|---|---|---|---|---|---|
| MW | MW | MW | MW/h | MW/h | ¥/MW | —— | —— | ¥/GJ | t/m$^3$ |
| 100 | 30 | 120 | 60 | 60 | 45 | 0.1 | 0.3 | 21 | 0.00234 |

| 100 | 30 | 120 | 60 | 60 | 45 | 0.1 | 0.3 | 21 | 0.00234 |

The remaining parameters are described below: ESS: $S_{min}$ = 32 MWh, $S_{max}$ = 200 MWh, $P^E_{D,max}$ = $P^E_{C,max}$ = 50 MW/h, $\eta^E$ = 0.95, $g_1 = g_2$ = 80 ¥/MW, $\lambda$ = 50 ¥/MW; NP unit: $P^N_{e,min}$ = 64 MW, $P^N_{e,max}$ = 100 MW, $P^N_{h,max}$ = 120 MW, $C_V$ = 0.3; HSS: $C_{max}$ = 160 MWh, $P^c_{h,max}$ = 40 MW; P2G unit and natural gas system: $\eta^{P2G}$ = 0.6, $P^{P2G}_{max}$ = 70 MW, $HHV$ = 36 MJ/m³, $\varepsilon$ = 3.6 MJ/kW·h; scheduling time: $T$ = 24, dispatch time $\Delta t$ = 1; Carbon trading model: $k_1$ = 40 ¥/t, $k_2$ = 120 ¥/t, $k_3$ = 200 ¥/t, $E_1$ = 1500 t, $E_2$ = 3000 t. To better analyze the role of carbon trading mechanism, the initial emission quota $E_f$ is 0. Based on previous research, the confidence level in this study was selected as 90%.

Based on the utilization of the NP units, this study sets three different modes for the simulation analysis:

Mode 1: TP and GC units bear electricity and heating loads, whereas NP units do not participate in the operation.

Mode 2: Based on mode 1, NP units are introduced to participate in the operation of the IES.

Mode 3: Based on mode 2, the NP units are transformed into NP cogeneration units and participate in the operation of the IES.

**5.2 Cost and carbon emission analysis of each operating mode**

In this study, the simulation of the above three modes was carried out, and the economic and environmental advantages brought about by the introduction of NP units and their nuclear heating renovation were verified through the analysis of operating costs and carbon emissions. The operating costs and carbon emissions of various modes are listed in Table 6. Note that the cost of heating renovation of the NP unit of the presented method was not considered in this study.

**Table 6. Comparison of operating costs and carbon emissions in different modes**

|  | mode 1 | mode 2 | mode 3 |
|---|---|---|---|
| TP operating cost/¥ | 445,125 | —— | —— |
| NP operating cost/¥ | —— | 1,200,000 | 1,200,000 |
| purchasing gas cost /¥ | 1,272,919 | 744,742 | 444,021 |
| carbon trading cost /¥ | 959,220 | 266,996 | 48,280 |
| total operating cost /¥ | 2,677,264 | 2,211,738 | **1,692,303** |
| total carbon emissions /t | 5,329 | 2,224 | **1,207** |

From Table 6, it can be seen that compared to mode 1, the carbon emissions of modes 2 and 3 are reduced by 3,105 t and 4,122 t, respectively. This illustrates that introducing NP units leads to a significant reduction in carbon emissions generated during the operation of the IES, which greatly enhances the environmental benefits. Compared to mode 1, the total operating costs of modes 2 and 3 were reduced by 455,526 ¥ and 984,961 ¥. However, if the carbon trading cost is not considered, mode 3 does not have an obvious economic advantage compared with that of mode 1, and even the operating cost of mode 2 is higher than that of mode 1. If the carbon trading mechanism is not introduced, the environmental benefit of carbon emission reduction cannot be transformed into an economic benefit for IES operators, which will be detrimental to the low-carbon transformation of IES. The operating costs and carbon emissions of mode 3 are significantly lower than those of mode 1, which proves that considering both NP units and carbon trading has clear environmental and economic benefits.

The operational status of each energy subsystem in different modes was compared and analyzed, as shown in Figs. 8-10.

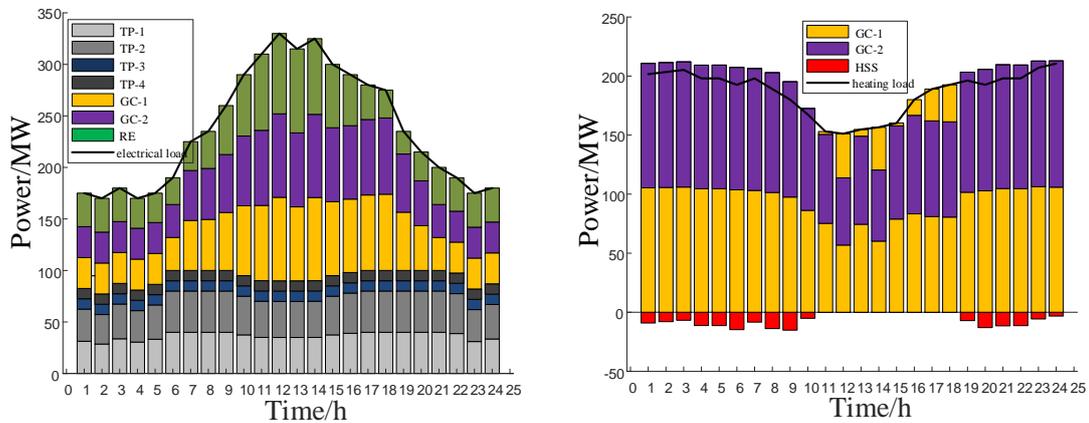

**Fig. 8 Operation of electrical and heating subsystems in mode 1**

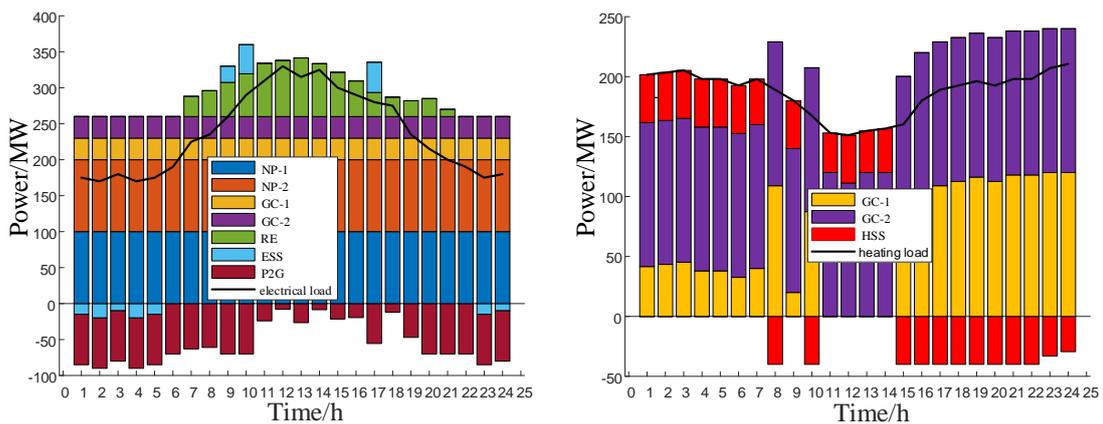

**Fig. 9 Operation of electrical and heating subsystems in mode 2**

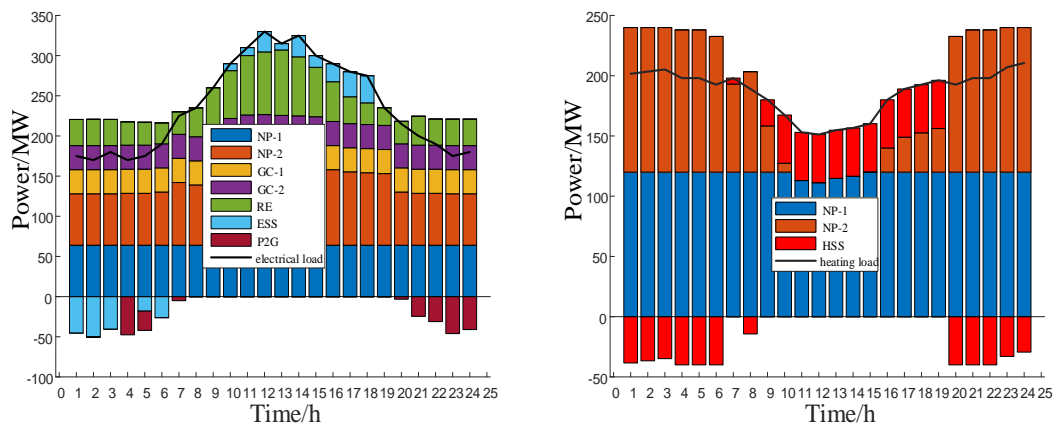

**Fig. 10 Operation of electrical and heating subsystems in mode 3**

As shown in Fig. 8, in mode 1, the electrical load is mainly borne by the TP and GC units, and the heating load is mainly borne by the GC unit. As can be seen from Fig. 9, NP units are introduced in mode 2. Compared with mode 1, TP units are out of operation at this time, and the electrical load is mainly borne by NP units, so the carbon emissions are significantly reduced. However, owing to the poor flexibility of NP units, many RE outputs have been reduced. The excess electricity is mainly converted into natural gas by a P2G unit, sent to the gas network, and then converted into electric energy and heat energy to supply the power and heating subsystem through the GC unit. Owing to the low efficiency of this conversion process, a large amount of energy loss occurs; thus, mode 2 has no obvious economic advantage compared with that of mode 1. As shown in Fig. 10, after the heating renovation of NP units in mode 3, both electrical and heating loads are mainly borne by NP units, while GC units operate in low-power mode and mainly provide spinning reserves for the system. Therefore, in mode 3, the operating costs and carbon emissions were significantly reduced.

**5.3 Flexibility improvement analysis of heating renovation of NP units**

To analyze the improvement in system flexibility brought about by the heating renovation of NP units, this study conducted a comparative analysis of the consumption of RE and the working situation of P2G in modes 2 and 3.

**5.3.1　Analysis of consumption of renewable energy**

A comparison of the RE consumption in modes 2 and 3 is shown in Fig. 11, and the specific circumstances of mode 3 are shown in Fig. 12.

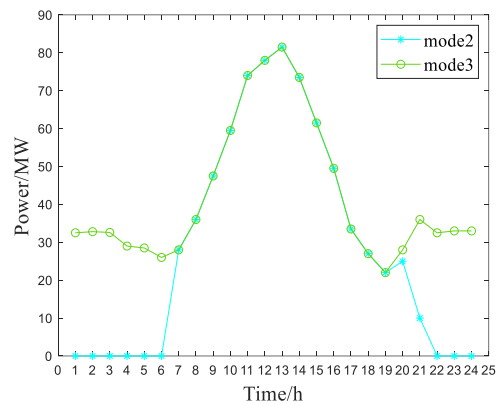

**Fig. 11 Consumption of renewable energy by different modes**

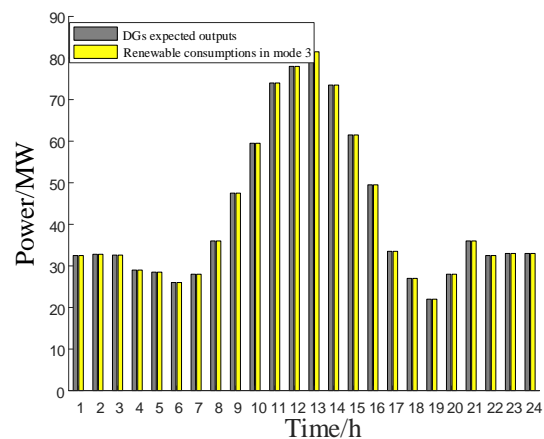

**Fig. 12 Renewable energy consumption in mode 3**

As shown in Figs. 11-12, the consumption of RE in the two modes is the same from 8:00 to-19:00, and complete consumption of RE is realized. This is because during 8:00 -19:00, the demand for electricity is high, and full utilization of RE can be realized. As can be seen from Fig. 11, during 0:00-7:00 and 19:00-24:00, a large amount of RE abandonment occurs in mode 2, and complete abandonment of RE output occurs during 0:00-6:00 and 22:00-24:00. This is because the electricity load level was low during these periods. At the same time, owing to the introduction of NP units, the system flexibility is poor, and to maintain the IES energy balance, a large amount of RE output was abandoned. In mode 3, the complete absorption of the RE is realized in any scheduling period. This is because the electrical and heating powers of the NP unit can be converted into each other within a certain range after heating renovation. When the electrical load is low in the early morning and at night, the heating power of the NP unit is appropriately increased, and the electrical power is reduced, thus increasing the absorption space of the RE. Therefore, heating the NP unit can improve the flexibility of its operation, increase the absorption space of the RE, and realize the complete absorption of the RE.

**5.3.2　Analysis of P2G unit operation**

In the P2G process, electrical energy is converted into natural gas and stored in the gas system. Although the conversion and storage of electrical energy can be achieved, the energy efficiency is not high, and there is significant energy loss during this process. Therefore, the utilization of P2G units, to a certain extent, is a reluctant action to achieve the internal energy balance of the IES due to poor system flexibility. The working conditions of the P2G units under different operating modes are shown in Fig. 13.

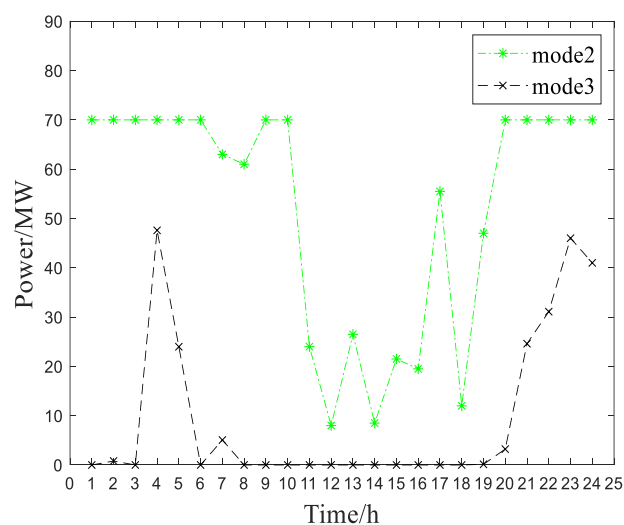

**Fig. 13 Operation of P2G units under different modes**

As shown in Fig. 13, in mode 2, the P2G unit is almost in a full-power operation state during early morning and night, and in a certain power operation mode during the daytime when the electrical load is high. However, the P2G units in mode 3 only run for a few hours in the early morning and night, and their operating power is much lower than that in mode 2. This is because in mode 2, the NP unit operates at the full electrical power state, and the P2G unit is forced to perform a significant amount of work to achieve electrical power balance. In mode 3, the NP units are transformed into cogeneration units, which can directly undertake heating demand, greatly improving their flexibility, and thus greatly reducing their dependence on the P2G unit to maintain power balance. It can be seen that the dependence on P2G units is greatly reduced owing to the improvement in system flexibility by NP unit heating renovation.

## 5.4 Analysis of joint operation of energy storage system
### 5.4.1　ESS operation analysis

Fig. 14 shows the charging and discharging of ESS in mode 3.

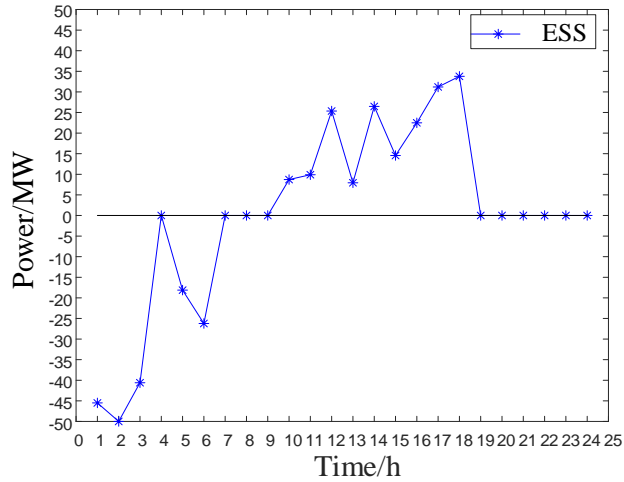

Fig. 14 Charging and discharging power of ESS

As can be seen from Fig. 14, the ESS is charged from 0:00-7:00, discharged from 9:00-19:00, and does not work in other periods. This is because the electric load is relatively low in the early morning, and the ESS is charged to store electricity. During the day, as the electric load increases, the ESS is discharged to maintain power balance. By charging and discharging at the appropriate time, the ESS actively participates in the dispatching operation of the IES, which realizes the peak shaving and valley filling of the electric load to a certain extent and improves the economy and flexibility of the system operation.

### 5.4.2　HSS operation analysis

The operation of the HSS in mode 3 is shown in Fig. 15.

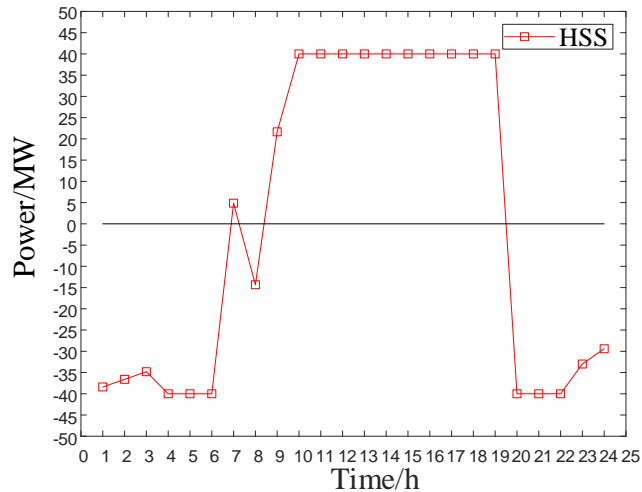

Fig. 15 Working condition of heat storage tank

As shown in Fig. 15, the HSS stores heat during 0:00-8:00 and 20:00-24:00 when the heat load demand is high, and releases heat from to 9:00-19:00 when the heating load demand is low. This is because the electrical load is low during 0:00-8:00 and 20:00-24:00, NP units can reduce the electrical power and increase the heating power to generate more heat and store it. During the daytime when the electrical load is high, part of the heating demand is provided by the HSS, so the heating power of the NP unit can be further reduced to improve the electrical power to meet the increasing electrical load demand. Through the storage and release of heat energy, the HSS can adjust the electrical power of the NP unit and improve its operational flexibility, which reflects the complementary advantages of multi-energy coupling.

It can be seen that joint operation of energy storage systems with different energy forms can realize complementary operation of various energy subsystems and improve the flexibility of system operation by decoupling thermoelectric power of cogeneration units.

### 5.5 Analysis of carbon trading mechanism

A carbon trading mechanism, taking $CO_2$ emission rights as the market trading object, can convert the low-carbon characteristics of NP unit into economic advantage, and the carbon trading mechanism in which the carbon trading price changes stepwise with carbon emissions can further expand this economic advantage of NP-IES, which is more conducive to the low carbon operation of IES. This section analyses the role of carbon trading mechanisms based on the impact of carbon trading price changes on carbon emissions and operating costs.

### 5.5.1 Analysis of the role of carbon trading price step changes

In order to analyze the impact brought by the step change of carbon trading price with carbon emission, the carbon trading cost and operation cost data of mode 1 and mode 3 under the fixed carbon trading price model and the stepped carbon price model were respectively compiled, and the results are shown in Table 7.

**Table 7. Comparison of carbon trading cost and operating cost in different price models**

|  | mode 1 | | mode 3 | |
|---|---|---|---|---|
|  | fixed price | stepped price | fixed price | stepped price |
| carbon trading cost/¥ | 657,161 | 959,220 | 144,842 | 48,280 |
| operating cost /¥ | 2,260,564 | 2,677,264 | 1,788,864 | 1,692,303 |

As can be seen from Table 7, by using the stepped carbon price model, the carbon trading cost of mode 1 with huge carbon emissions is further increased, while the carbon trading cost of mode 3 with little carbon emissions is significantly reduced. Regarding operating costs, the economic advantage of mode 3 over mode 1 has been expanded from 471,700¥ to 984,961¥. These results suggest the stepped carbon price model manages to expand the economic advantages of the NP-IES with low-carbon emission characteristics.

### 5.5.2  Impact of carbon trading price on carbon emissions

This section analyzes the impact of the change in the carbon trading price on carbon emissions in different modes. In modes 2 and 3, since the NP units bears most of the load, the impact of the carbon trading prices on carbon emissions is not obvious. In term of mode 1, the change of carbon emission with carbon trading prices is illustrated in Fig. 16.

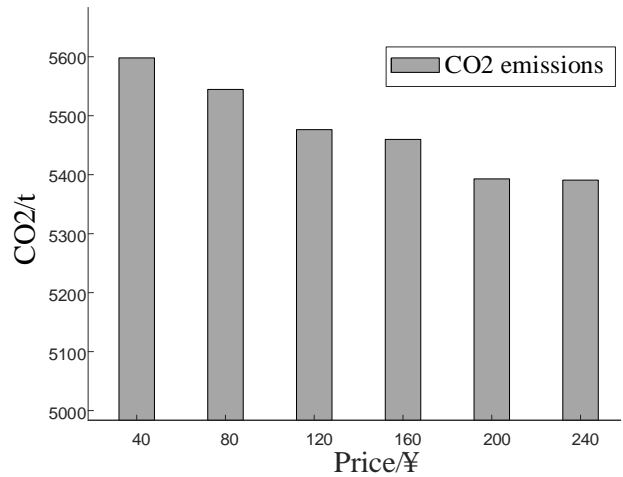

Fig. 16 Change of carbon emission with carbon trading prices

As can be seen from Fig. 16, when the carbon trading price is lower than 200 ¥/t, the carbon emissions decrease with a continuous increase in the carbon trading price. When the carbon trading price exceeds 200 ¥/t, the carbon emissions do not change. This is because when the carbon trading price is low, with the increase in carbon trading price, TP units with lower power generation cost and higher carbon emission intensity are gradually replaced by GC units that have higher power generation cost and lower carbon emission intensity; thus, the carbon emissions continue to decline. When the carbon trading price exceeds 200 ¥/t, the GC unit is already in the full power operation state, and it cannot replace more output of the TP units, so the carbon emissions do not change. It can be seen that an appropriate carbon trading mechanism, in which the carbon trading price can change in steps with the carbon emissions, is conducive to the priority outputs of low-carbon emission units, thus promoting the reduction of carbon emissions. However, the carbon trading mechanism cannot change the phenomenon of coal and natural gas combustion to supply the base load of the IES, so its independent operation has a limited role in promoting carbon emission reduction.

### 5.5.3 Impact of carbon trading price on operating cost

The change in the total operating cost with the carbon trading prices under the different operating modes is shown in Fig. 17.

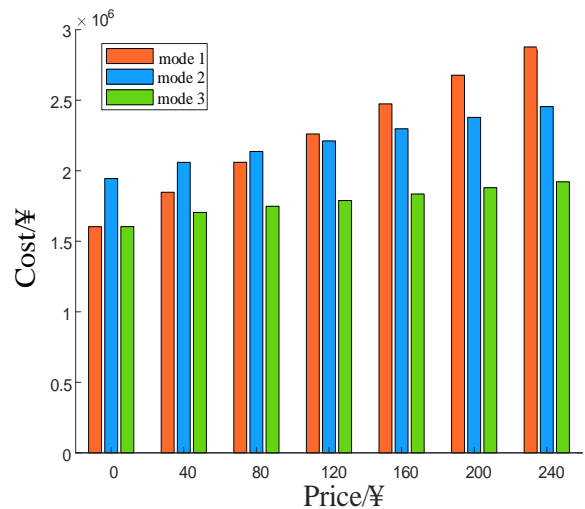

**Fig. 17 Change of total operating cost with carbon trading prices**

Fig. 17 shows that the total operating cost of the three modes increases with the increase in carbon trading price. When the carbon trading cost is not calculated, the operating cost of mode 2 is greater than that of mode 1, because the reduced flexibility brought to the system by introducing NP units alone increases the operating cost. The operating cost of mode 3 is basically the same as that of mode 1, without obvious advantages. With the increase in carbon trading price, the advantages of mode 2 and mode 3 gradually become more apparent. When the carbon trading price is 105 ¥/t, the operating cost of mode 2 begins to be lower than that of mode 1, and mode 3 has an even greater advantage over mode 1. The reason for this is that the carbon emission in mode 1 is relatively large, while those in modes 2 and 3 are relatively small. The economic advantage brought by the low-carbon characteristics of NP units is further highlighted by using our designed stepped carbon price model. With the improvement in public awareness of environmental protection and the increasingly urgent requirements for energy conservation and emission reduction, the carbon trading price will continue to rise in the future, making the low-carbon economic advantages of the IES with NP units more and more obvious.

### 5.5.4 Analysis of the variation in outputs of TP and GC units with carbon trading prices

The variation in outputs of TP and GC units with carbon trading prices is shown in Fig. 18.

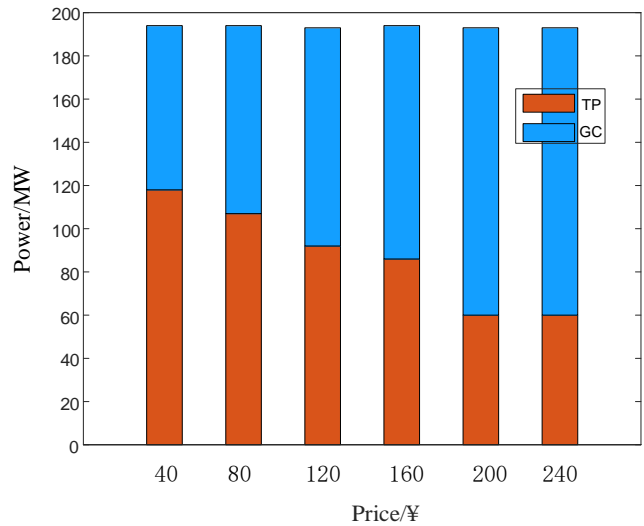

**Fig. 18 Variation in outputs of TP and GC units with carbon trading prices**

Fig. 18 shows that as the carbon trading price increases, the power output of TP units gradually decreases, while that of GC units gradually increases. This is because TP units have low operating cost and high carbon emission intensity; on the contrary, GC units have high operating cost and low carbon emission intensity. Without considering carbon emission cost, TP units have the price advantage and are given priority in power generation. As the carbon trading price gradually increases, GC units with low carbon emission intensity gradually start to have economic advantages and gradually replace the output of TP units. This shows that the introduction of carbon trading in IES is able to promote priority power generation for units with low-carbon emission intensity in the system, which has a positive effect on the reduction in carbon emissions. After the carbon trading price exceeds 200 ¥, the GC unit reaches its maximum power. In this situation, the ratio of the powers of GC and TP units no longer changes.

From the above analysis, it can be seen that carbon trading can transform the low-carbon characteristics of NP units into economic advantages, and the stepped carbon price model can further expand this advantage, thereby contributing to the low-carbon development of the IES.

**5.6 Comparison of solution algorithms**

To demonstrate the superiority of the method proposed in this paper, we compare it with some scenario-based methods such as scenario approach (SA) and sample average approximation (SAA), which are commonly used to solve chance-constrained models. Regarding the setting of related parameters, we mainly refer to (Doluweera et al., 2020) and (Pagnoncelli et al., 2009). Considering inherent randomness of scenario-based methods, the average result of 20 independent runs is used as the final result. The comparison results are shown in Table 8.

**Table 8. Comparison of the results of different solution algorithms**

|  | Operating cost/¥ | Carbon emission/t | Calculation time/s |
|---|---|---|---|
| SA | 1,703,457 | 1,339 | 38.4 |
| SAA | 1,722,473 | 1,418 | 16.1 |
| Proposed method | **1,692,303** | **1,207** | **0.94** |

As indicated in Table 8, the proposed method is superior to the SA and SAA in terms of optimization result and calculation time. In term of operating costs, the cost of our proposal is respectively lower than those of the SA and SAA by 11,154 ¥ and 30,170 ¥; as for carbon emissions, the emission of the proposed method is lower than those of the two comparative algorithms by 132 t and 211 t; regarding calculation times, the time of ours is lower than those of other alternatives by 37.46 s and 15.16 s. Therefore, it can be concluded that our method is the best one in this study for handling chance-constrained optimization.

**6  Conclusions**

To promote the low-carbon operation of an IES, this paper introduces a carbon trading mechanism and NP units into an IES, and constructs a new low-carbon operation and scheduling model. Based on the simulation results, the following conclusions can be drawn:

1) In the newly constructed low-carbon IES, the carbon trading mechanism and the nuclear power units operate in coordination with each other, which greatly reduces the $CO_2$ emissions and overall operating costs of the entire system and provides good economic and environmental benefits.

2) The operational flexibility of the nuclear power unit and the entire system was improved after the renovation of the nuclear power unit. In addition, electric energy storage systems, heat storage systems, and auxiliary equipment such as P2G units operate in coordination and complementarity, realizing reasonable energy transfer and conversion, alleviating the spatio-temporal mismatch between energy supply and demand, improving the flexibility of system operation, and expanding the consumption space of RE.

3) The proposed model solution based on sequence operation theory can transform the hard-to-solve chance-constrained programming model into an easily solvable mixed-integer programming model; thus, an optimal scheduling scheme can be obtained quickly and accurately.

In this study, the modeling of the thermoelectric characteristics of nuclear power cogeneration units was simplified to some extent. In the actual model, there is a certain amount of energy loss in the working process of the steam turbine, and the relationship between the heating supply and electricity loss is not typically linear. In addition, this study did not consider the cost of heating renovation of the NP unit, and more realistic scenarios need to consider this cost. Leveraging flexible loads and demand response to improve system flexibility will be the exploration direction of our future work. This study uses probability density functions to depict renewable power, and it is interesting to employ generative adversarial network-based renewable scenario generation to more realistically model the renewable uncertainties (Li et al., 2018b).